\documentclass[12pt,a4paper]{article}
\usepackage{graphicx}
\begin{document}
\begin{center}
 {\bf Non-Markovian effects in quantum system: an exact stochastic mean-field treatment.}

 \medskip
G. Hupin and D. Lacroix$^{1}$

 \smallskip
{\it
 $^{1}$GANIL-France \\
}
\end{center}

\bigskip
\begin{abstract}
A quantum Monte-Carlo is proposed to describe fusion/fission processes when fluctuation and dissipation, with memory effects, 
are important. The new theory is illustrated for systems with inverted harmonic potentials coupled to a heat-bath.
\end{abstract}

To understand fusion and fission processes, few relevant degrees of freedom (DoF) that contain important
aspects of the nuclear dynamics: relative distance, mass asymmetry, deformation ..., are selected. To properly treat these mechanisms, 
relevant DoF should be seen as an Open Quantum System (OQS) coupled to the surrounding complex environment of irrelevant 
DoF (called hereafter environment). Most recent advances in OQS should be used in these cases, since close to the barrier 
dissipation, fluctuation and memory (also called Non-Markovian) effects are much involved in the dynamics. Recently, we have proposed a new 
method to describe a system (S) + environment (E) with a total Hamiltonian given by
\begin{eqnarray}
H = h_S + h_E + Q \otimes B, 
\end{eqnarray}   
where $h_S$ and $h_E$ denote the system and environment Hamiltonians respectively while $Q \otimes B$ is responsible for the coupling. 
The system density evolution, denoted by $\rho_S$, can be exactly replaced by a stochastic integro-differential master equation written as:
\begin{eqnarray} 
d \rho_S =&& \frac{dt}{i\hbar} \left[h_S , \rho_S \right] + dt [Q,\rho_S] \int^t_0 ds D(t-s) \left\langle  Q(s) \right\rangle_S \nonumber \\   
&+& d \xi(t) [Q,\rho_S] + d\eta(t) \{Q - \left\langle  Q \right\rangle_S , \rho_S \} 
\label{eq:stocmfsimple} 
\end{eqnarray}
with
\begin{eqnarray}
d\xi(t) &=&  dt \int^t_0 D_1(t-s) dv_E(s) - dt \int^t_0 D(t-s) du_E(s) -idv_S(t) ,~d\eta(t) = du_S (t). \nonumber
\end{eqnarray}
Noises $du_{S/E}$ and $dv_{S/E}$ are Gaussian stochastic variables 
with zero mean values and variances equal to:
\begin{eqnarray}
\begin{array}{ccc}
\overline{du_Sdu_E} &=& \overline{dv_Sdv_E} = \frac{dt}{2\hbar},~~~~
\overline{du_Sdv_E} = \overline{dv_Sdu_E} = 0. 
\end{array}
\label{eq:noiseusue}
\end{eqnarray}
In Eq. (\ref{eq:stocmfsimple}), $D$ and $D_1$ denote time memory functions characterizing the environment [2]. 

\begin{figure}[t!]
\begin{center}
\includegraphics[width = 0.9\textwidth]{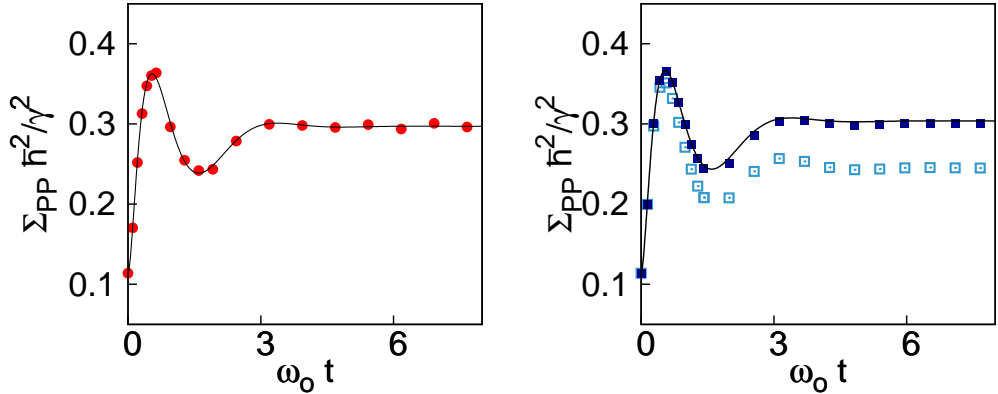}
\end{center}
\caption{Evolution of $\Sigma_{PP}$ obtained with quantum Monte-Carlo (QMC, left) and TCL (right) formalisms are shown as a function of time and compared to the  exact evolution (solid line). In the TCL case, the second order (TCL2, open squares) or fourth order (TCL4, filled squares) in the coupling constant are displayed. Parameters of the simulation are $k_b T = \hbar \omega_0$, $\eta =0,5 \hbar \omega_0$, $\hbar \Omega = 5 \hbar\omega_0$ and $\hbar \omega_0 = 14 MeV$. }
\label{fig:SMEsim}
\end{figure}

To illustrate the method, we consider the Caldeira-Leggett (CL) model which can be analytically solved and consists of a harmonic system coupled to a bath of oscillators [3]  with: 
\begin{eqnarray}
h_S = h_c + \frac{P^2}{2M} + \varepsilon \frac{1}{2}M\omega_0^2 Q^2,~~
h_E = \sum_{n}\left(\frac{p^2_n }{2m_n} + \frac{1}{2}m_n \omega^2_n x^2_n\right)
\end{eqnarray}
while $B \equiv -\sum_n \kappa_n x_n$ [1]. Here, $h_c = Q^2 \sum_n \kappa^2_n /(2m_n \omega^2_n)$ is the counter-term that 
insures that the physical frequency is $\omega_0$. The thermal behavior of the environment is completely defined by its spectral density [1,5]  $\displaystyle J(\omega)\equiv \sum_n \frac{ \kappa^2_{n}} {2m_n \omega_n} \delta(\omega - \omega_n)$. In the following, a Drude spectral density [2]
\begin{eqnarray}
J(\omega) &=& \frac{2 M \eta}{\pi} \omega \frac{\Omega^2}{\omega^2+ \Omega^2},
\end{eqnarray}
is considered where $M$ is the nucleon mass, and $\eta$ denotes the strength of the coupling.
In this model, the master equation (\ref{eq:stocmfsimple}) can be solved by following 
the stochastic evolution of first and second moments of $P$ and $Q$.
The QMC efficiency 
has been systematically investigated for various temperatures and coupling strengths in the harmonic case ($\varepsilon = 1$) [7].
In all cases, averaged evolutions could almost not be distinguished from the exact evolution. This is illustrated 
in figure \ref{fig:SMEsim} where the average evolution of momentum dispersion, denoted by $\Sigma_{PP}$ (left, filled circle), 
is displayed as a function of time and compared to the exact solution (solid line). This approach is as accurate as the fourth order in coupling (filled squares) of the Time ConvolutionLess (TCL) method (right). This expansion method consists in a time local projective method [1] and leads to master equation with time-dependent transport coefficients:
\begin{eqnarray}
\hbar \frac{d}{dt}\rho_S(t) =\hspace{-1.5mm}&-&\hspace{-1.5mm} i[H_S,\rho_S(t)] - \frac{i}{2} \Delta(t) [Q , \{Q,\rho_S(t) \}]  - 2 i \lambda(t)[Q,\{P,\rho_S(t)\}] \nonumber\\ \hspace{-1.5mm}& -&\hspace{-1.5mm} \frac{D_{PP}(t)}{\hbar}[Q,[Q,\rho_S(t)]] + 2 \frac{D_{PQ}(t)}{\hbar}[Q,[P,\rho_S(t)]]. 
\end{eqnarray}
This method, when second order in the coupling constant is retained, is the preferred tool to treat non-Markovian effects in nuclear physics.   

The inverted harmonic oscillator ($\varepsilon=-1$) is the first step towards realistic applications to fusion and fission. In these physical situations, the passing probability, i.e. the probability to pass through the barrier, is of particular importance. It is usually defined as:
\begin{eqnarray}
P(+\infty) &=&  \lim_{t\to + \infty}\frac{1}{2} {\rm erfc} \left( - \frac{|q(t)|}{\sqrt{2 \Sigma_{qq} (t)}} \right), \label{eq:proba}
\end{eqnarray}
where $ q(t)$ and $\Sigma_{qq} (t)$ denote the expectation value and second moment 
of $Q$ deduced from the considered theory. We have systematically investigated 
the difference between the estimated passing probability and the exact one using the parameter $\Delta P/P$:
\begin{eqnarray}
\frac{\Delta P}{P} &\equiv & \frac{P(+\infty) - P_{\rm ex}(+\infty)}{P_{\rm ex}(+\infty)}
\end{eqnarray}
where $P(+\infty)$ and $P_{ex}(+\infty)$ denote the results of the specific calculation considered and the exact one respectively.

\begin{figure}[t!]
\begin{center}
\includegraphics[width = 0.9\textwidth]{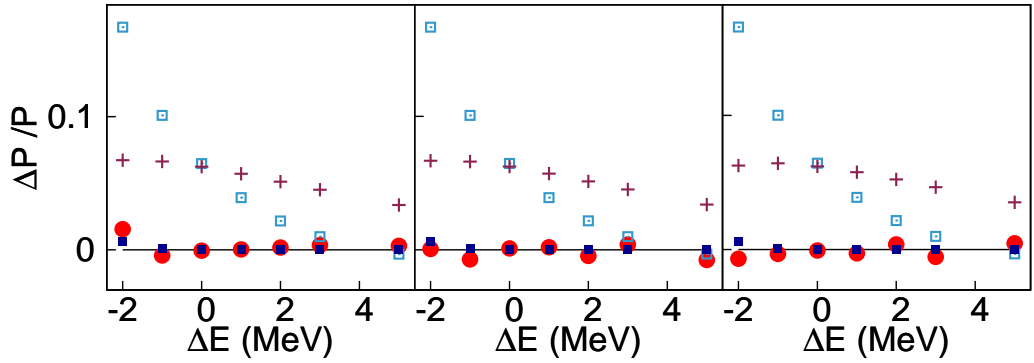}
\end{center}
\caption{Evolution of $\Delta P / P$ as a function of $\Delta E$ calculated using 
quantum Monte-Carlo (filled circles), TCL2 (open squares), TCL4 (filled squares) and Markovian 
approximation (open crosses) for different temperatures: from left to right $T= 5\hbar \omega$, $\hbar \omega$ and $0,1\hbar\omega$ (adapted 
from [7]).}
\label{fig:prob}
\end{figure}

Figure \ref{fig:prob} presents the evolution of $\Delta P/P$ as a function of $\Delta E$, the total system energy minus the barrier hight, obtained for different temperatures for the quantum Monte-Carlo (filled circles), second order of the TCL method (TCL2, open squares), 
TCL4 (filled squares) cases. In this figure, the Markovian approximation is also shown by open crosses.

The TCL4 and the quantum Monte-Carlo methods are in perfect agreement with the exact solution for any input parameters. 
Small differences sometimes observed between the stochastic approach and the exact value come from the limited number of trajectories 
used to obtain figure \ref{fig:prob}. At low $\Delta E$, TCL2 turns out to be a rather poor approximation which confirms the difficulty of treating non-Markovian effects below the barrier. It is worth finally to mention 
that below the barrier, the Markovian limit gives an even better result than the TCL2 case. \\

Quantum Monte-Carlo method is a competitive theory to simulate non-Markovian effects in Open Quantum Systems. The present application to inverted parabola is the first step towards realistic treatment of fusion/fission.

\bigskip
[1] H .P. Breuer and F. Petruccione, "The Theory of Open Quantum Systems", (2002), Oxford University Press. \\
\noindent
[2] D. Lacroix, Phys. Rev. {\bf E77}, 041126 (2008).\\
\noindent
[3] A. O. Caldeira, A. J. Leggett, Ann. of Phys. {\bf 149}, 374 (1983).\\
\noindent
[4] D. Lacroix and G. Hupin, Proceedings of " FUSION08: New Aspects of Heavy Ion Collisions near the Coulomb Barrier", September 22-26, 2008, Chicago, USA, arXiv:0812.3650.\\
\noindent
[5] J. Leggett, S. Chakravarty, A. T. Dorsey, Matthew P. A. Fisher, Anupam Garg, and W. Zwerger, Rev. Mod. Phys. {\bf 59}, 1 (1987).\\
\noindent
[6] H. Goutte, J. F. Berger, P. Casoli and D. Gogny, Phys. Rev. {\bf C71}, 024316 (2005).

[7] G. Hupin and D. Lacroix, {\it in preparation}. \\

\end{document}